\documentclass[12pt]{article}

\usepackage{graphics}

\def\Red  {}
\def\Black{}
\def\Blue {}
\begin{document}
\title{\Red \bf
        Approximate Analytic Solutions \\of  RG  Equations
         for Yukawa\\ and Soft Couplings in SUSY Models
        \Black
\thanks{Published in  Eur. Phys. J. ~{\bfseries C 13}, 671-679 (2000); hep-ph/9906256}
       }
\author{Sorin Codoban\footnote{\sf codoban@thsun1.jinr.ru},
        Dmitri I. Kazakov\footnote{\sf kazakovd@thsun1.jinr.ru}\\[3mm]
{\em Bogoliubov Laboratory of Theoretical Physics, JINR,}\\
{\em 141 980 Dubna, Moscow Region, Russian Federation}}
\date{}
%

% 11.10.Hi Renormalization group evolution of parameters
% 11.30.Pb Supersymmetry

%\twocolumn[\begin{@twocolumnfalse}
\maketitle
\begin{abstract}
\Blue
We present simple analytical formulae which describe solutions of
the RG equations for Yukawa couplings in SUSY gauge theories with the
accuracy of a few per cent. Performing the Grassmannian expansion in these
solutions, one finds those for all the soft couplings and masses. The
solutions clearly exhibit the fixed point behaviour which can be calculated
analytically. A comparison with numerical solutions is made.\Black
\end{abstract}
\vspace*{0.5cm}
%\end{@twocolumnfalse}]

\section{Introduction}
\label{intro}
The renormalization group equations (RGEs) for the rigid couplings
and soft parameters in SUSY gauge theories play a crucial role in
applications. Actually, all predictions of the MSSM are based on
solutions to these equations in  leading and next-to-leading
orders \cite{SUSY}. Typically, one has three gauge couplings, one or
three Yukawa couplings (for the case of low or high $\tan\beta$,
respectively) and a set of soft couplings. In leading order,
solutions to the RGE for the gauge couplings are simple; however,
already for Yukawa couplings, they are known in an analytical form
only for the low $\tan\beta$ case, where  only the top
coupling is left. Moreover, even in this case solutions for the soft
terms look rather cumbersome and difficult to
explore~\cite{Ibanez}.

In a recent paper~\cite{K}, it has been shown  that solutions to
the RGE for the soft couplings follow from those for the rigid
ones in a   straightforward
way.\footnote{Here we follow the approach advocated in Ref.\cite{AKK}.\\
\indent A similar method but in somewhat different way
        has been also presented in Refs.\cite{JJ,GR}.}
One takes the solution for the
rigid coupling (gauge or Yukawa), substitute instead of the
initial conditions their modified expressions
%~\cite{AKK}
\begin{eqnarray}
\alpha_i &\Rightarrow&
            \tilde{\alpha}_i=\alpha_i(1+M_i \eta+\bar M_i
            \bar{\eta}+2M_i\bar M_i \eta \bar{\eta}),  \label{g}
\\ Y_k &\Rightarrow &
          \tilde{Y}_k= Y_k(1-A_k\eta -\bar{A}_k\bar{\eta}
           +A_k\bar{A}_k\eta\bar{\eta} +\Sigma_k\eta\bar{\eta}). \label{Y}
\end{eqnarray}
where $\eta=\theta^2, \bar{\eta}=\bar{\theta}^2$,
$\theta $ and $\bar \theta$ are the Grassmannian parameters,
and expand over these parameters. This
gives the solution to the RGEs for the soft couplings.

Hereafter the following notation is used:
\begin{equation}
  \alpha_i \equiv \frac{g_i^2}{16\pi^2} ,  \ \
 Y_k \equiv \frac{y_k^2}{16\pi^2}\, , \ \
 \Sigma_k = \sum_{j=1}^{3}m^2_j \, .
\end{equation}
where $g_i$ and $y_k$ are the gauge and Yukawa couplings, respectively,
 and $m_j^2$ are the soft masses associated with each scalar field.

This procedure, however, assumes that one knows solutions to the RGE for the
rigid couplings in the analytic form.  For instance, in the case of the MSSM in
the low $\tan\beta$ regime this allows one to get  solutions for the soft
couplings and masses simpler than those known in the literature ~(see~\cite{K}).
At the same time, in many cases such solutions are unknown. Actual examples
are the MSSM with high $\tan\beta$ and NMSSM. One is bound to solve the
RGEs numerically when the number of coupled equations increases dramatically
with the soft terms being included.

Below we propose simple analytical formulae which give an approximate
solution to the RGE for Yukawa couplings in an arbitrary SUSY theory with the
accuracy of a few per cent. Performing the Grassmannian expansion in these
approximate solutions one can get  those for the soft couplings in a
straightforward way. As an illustration we consider the MSSM in the high
$\tan\beta$ regime.

  One can immediately see that approximate solutions obtained in this way possess
infrared quasi-fixed points~\cite{Hill} which can be found analytically.
 They appear in the limit when the initial values of the Yukawa couplings are much
larger than those for the    gauge ones. Then, one can analytically trace how
the initial conditions for the soft terms disappear from their solutions
in the above~ mentioned limit.

The paper is organized as follows.  In Sect. 2, we consider
  the MSSM in the low $\tan\beta$ regime, where all solutions are known
  analytically and describe briefly the Grassmannian expansion.
In Sect. 3, we present our approximate solutions for the
   Yukawa couplings and obtain those for the soft terms. We also present
   numerical illustration and compare  approximate solutions with the
   numerical ones.
The fixed point behaviour is discussed. Section 4 contains our conclusions.
The explicit formulae for the soft couplings and masses are given in Appendices.
\section{The MSSM: exact solutions in the low $\tan\beta$ case}\label{sect1}
  Consider the MSSM in the low $\tan\beta$ regime.  One has three gauge and one
  Yukawa coupling. The one-loop RG equations are
\begin{eqnarray}
\dot{\alpha}_i&=&-b_i\alpha^2_i, \ \ \ \  b_i=(\frac{33}{5},1,-3),
\ \ i=1,2,3. \\
\dot{Y}_t&=&Y_t\left(\frac{16}{3}\alpha_3+3\alpha_2+\frac{13}{15}\alpha_1-6Y_t\right).
\end{eqnarray}
with the initial conditions: $\alpha_i(0)=\alpha_0, \ Y_t(0)=Y_0$,
and $t=\ln(M_X^2/Q^2)$.
 Their solutions are given by~\cite{Ibanez}
\begin{equation}
\alpha_i(t)=\frac{\alpha_0}{1+b_i\alpha_0t}\, , \ \ \
     Y_t(t)=\frac{Y_0E(t)}{1+6Y_0F(t)}\, ,
 \label{sol}
\end{equation}
where
\begin{eqnarray}
E(t)&=&\prod_i(1+b_i\alpha_0t)^{\displaystyle c_i/b_i} , \ \ \
 c_i=(\frac{13}{15},3,\frac{16}{3}), \label{e}\\
F(t)&=&\int^t_0 E(t')dt'. \label{f}
\end{eqnarray}
   To get solutions for the soft terms, it is enough to perform the
substitution $\alpha \to \tilde{\alpha}$ and $Y\to \tilde{Y}$ for the initial
conditions in (\ref{sol}) and expand over $\eta $ and $\bar \eta$.
   Expanding the gauge coupling in (\ref{sol}) up to $\eta$ one has
(hereafter we assume $M_{i0}=m_{1/2}$)
\begin{equation}
M_i(t)=\frac{m_{1/2}}{1+b_i\alpha_0t}.
\end{equation}
  Performing the same expansion for the Yukawa coupling and using
the relations
\begin{equation}\label{tayl}
\left. \frac{d\tilde{E}}{d\eta}\right|_\eta=m_{1/2}t\frac{dE}{dt}, \ \ \
\left.\frac{d\tilde{F}}{d\eta}\right|_\eta=m_{1/2}(tE-F).
\end{equation}
 one finds the well-known expression~\cite{Ibanez}
\begin{equation}
A_t(t)=\frac{A_0}{1+6Y_0F}+m_{1/2}\left( \frac{t}{E}\frac{dE}{dt}-
\frac{6Y_0(tE-F)}{1+6Y_0F}\right)\!. \label{a}
\end{equation}
 To get the solution for the $\Sigma$ term, one has to make
expansion over $\eta$ and $\bar \eta$. This can be done with the
help of the following relations:
\begin{equation}
 \left.\frac{d^2\tilde{E}}{d\eta
d\bar \eta}\right|_{\eta,\bar \eta}=
m_{1/2}^2\frac{d}{dt}\left(t^2\frac{dE}{dt}\right), \ \ \
\left.\frac{d^2\tilde{F}}{d\eta d\bar \eta}\right|_{\eta,\bar \eta}
=m_{1/2}^2t^2\frac{dE}{dt}\, .
\end{equation}
 As a result one has ~\cite{K}
\begin{eqnarray}
\Sigma_{t}(t)\!\!\!\!\!&=&\!\!\!\!\!
  \frac{\Sigma_0\!-A_0^2}{1\!+6Y_0F}+\frac{(A_0\!-m_{1/2}6Y_0(tE\!-F))^2}{(1+6Y_0F)^2}
 +m_{1/2}^2\left[\frac{d}{dt}\left(\frac{t^2}{E}\frac{dE}{dt}\right)
        \!\!-\!\frac{6Y_0}{1\!+6Y_0F}t^2\frac{dE}{dt}\right]\!.\, \label{si}
\end{eqnarray}
which is much simpler than what one finds in the literature~\cite{Ibanez},
though coinciding with it after some cumbersome algebra.

 One can also write down  solutions for the individual masses
using the Grassmannian expansion of those for the corresponding
superfield propagators.  For the first two generations one has
\begin{eqnarray}
m^2_{Q_L}&=&m_0^2+\frac 12 m_{1/2}^2\left(\frac{16}{3}f_3+3f_2+\frac{1}{15}f_1\right),\label{mq} \\
m^2_{U_R}&=&m_0^2+\frac 12 m_{1/2}^2\left(\frac{16}{3}f_3+\frac{16}{15}f_1\right),\\
m^2_{D_R}&=&m_0^2+\frac 12 m_{1/2}^2\left(\frac{16}{3}f_3+\frac{4}{15}f_1\right),\\
m^2_{H_1}&=&m_0^2+\frac 12 m_{1/2}^2\left(3f_2+\frac{3}{5}f_1\right),\\
m^2_{L_L}&=&m_0^2+\frac 12 m_{1/2}^2\left(3f_2+\frac{3}{15}f_1\right),\\
m^2_{E_R}&=&m_0^2+\frac 12 m_{1/2}^2\left(\frac{12}{5}f_1\right), \label{me}
\end{eqnarray}
  where
\begin{equation}
f_i=\frac{1}{b_i}\left(1-\frac{1}{(1+b_i\alpha_0t)^2}\right).
\end{equation}
 The third generation masses get a contribution from the top Yukawa coupling
\begin{eqnarray}
m^2_{b_R}&=&m^2_{D_R}, \\ m^2_{b_L}&=&m^2_{D_L}+\Delta/6, \\
 m^2_{t_R}&=&m^2_{U_R}+\Delta/3, \\
m^2_{t_L}&=&m^2_{U_L}+\Delta/6, \\ m^2_{H_2}&=&m^2_{H_1}+\Delta/2,
\end{eqnarray}
  where $\Delta$ is related to $\Sigma_t$ (\ref{si}) by
\begin{eqnarray*}
 \Delta &=& \Sigma_t-\Sigma_0
            -m_{1/2}^2\left[\frac{d}{dt}\left(\frac{t^2}{E}\frac{dE}{dt}\right)\right] \\
 &=& \frac{\Sigma_0-A_0^2}{1+6Y_0F}+\frac{(A_0-m_{1/2}6Y_0(tE-F))^2}{(1+6Y_0F)^2}
 -m_{1/2}^2\frac{6Y_0}{1+6Y_0F}t^2\frac{dE}{dt}- \Sigma_0.
\end{eqnarray*}
  With  analytic solutions (\ref{sol},\ref{a},\ref{si}) one can analyze
 asymptotic and, in particular, find infrared quasi-fixed
points~\cite{Hill} which correspond to $Y_0 \to \infty$
\begin{eqnarray}
Y_t^{FP}\!\! &=&\! \frac{E}{6F},  \label{Yf} \\
A_t^{FP}\!\! &=&\! m_{1/2}\left(\frac{t}{E}\frac{dE}{dt}-\frac{tE-F}{F}\right), \label{Af} \\
\Sigma_t^{FP}\!\! &=&\!m_{1/2}^2
      \left[\left(\frac{tE-F}{F}\right)^2\!\!+\!\frac{d}{dt}
      \left(\frac{t^2}{E}\frac{dE}{dt}\right)\!-\!\frac{t^2}{F}\frac{dE}{dt}\right]\!. \label{Sf}
\end{eqnarray}
  One can clearly see that the dependence on $Y_0, A_0$ and $\Sigma_0$ disappears
from (\ref{Yf})-(\ref{Sf}).
  Some residual dependence on $m_0^2$ is left for the soft masses and partially
cancels with that of $\Delta$.

 Below we demonstrate  how the same procedure works in the case of approximate
solutions. As a realistic example we take the MSSM in the high $\tan\beta$ regime.
   \section{The MSSM: approximate solutions in high $\tan\beta$ case}\label{sect2}
The one-loop RGE for the Yukawa couplings in this case look like
\begin{eqnarray*}\label{yeq}
   \dot{Y}_t &=& Y_t\left(\frac{16}3\alpha _3+3\alpha _2+\frac{13}{15}\alpha _1-6Y_t-Y_b\right) ,\\
   \dot{Y}_b &=& Y_b\left(\frac{16}3\alpha _3+3\alpha _2+\frac 7{15}\alpha _1-Y_t-6Y_b-Y_\tau \right), \\
\dot{Y}_\tau &=& Y_\tau \left(3\alpha _2+\frac 95\alpha _1-3Y_b-4Y_\tau \right).
\end{eqnarray*}
  Since the exact solution is absent and might be too cumbersome, we look for
an approximate one in a simple form similar to that of (\ref{sol}).
  \subsection{The choice of approximate solution}\label{analyt}
 In choosing approximate solutions we follow the idea of \cite{CW}
where an approximate solution for $Y_t$ and $Y_b$ ignoring $Y_\tau$ has been
proposed. Our suggestion is to consider separate brackets for each
propagator entering into the Yukawa vertex. Then, one has the following
expressions for the Yukawa couplings:
\begin{eqnarray*}
Y_t&=&\frac{\displaystyle Y_{t0}E_t}{\displaystyle [1+A(Y_{t0}F_t+Y_{b0}F_b)]^{1/A}
   [1+2BY_{t0}F_t]^{1/B} [1+3CY_{t0}F_t]^{1/C} },\quad  \frac 1A \!+\! \frac 1B \!+\! \frac 1C\!=\!1    \\
   Y_b&=&\frac{\displaystyle Y_{b0}E_b}{\displaystyle [1+A(Y_{t0}F_t+Y_{b0}F_b)]^{1/A}
   [1+2BY_{b0}F_b]^{1/B} [1+C(3Y_{b0}F_b+Y_{\tau 0}F_\tau )]^{1/C} },\\
 Y_\tau &=& \frac{\displaystyle Y_{\tau 0}E_\tau }
    {\displaystyle [1+A^{\prime }Y_{\tau 0}F_\tau ]^{1/A^{\prime }}
                   [1+2B^{\prime }Y_{\tau 0}F_\tau]^{1/B^{\prime}}
                   [1+C(3Y_{b0}F_b+Y_{\tau 0}F_\tau )]^{1/C}},\quad
                  \frac1{A^{\prime }}\!+\!\frac 1{B^{\prime }}\!+\!\frac 1C\!=\!1
\end{eqnarray*}
where the brackets correspond to the $Q,U,H_2$, $Q,D,H_1$ and $L,E,H_1$
propagators, respectively. Here $E_t$ and $F_t$ are given by
(\ref{e}) and (\ref{f}) and $E_b$ and $E_\tau$ have the same form but with
$c_i^{(b)}=(7/15,3,16/3)$ and $c_i^{(\tau)}=(9/5,3,0)$, respectively.

 The brackets are organized so that  they reproduce the contributions
of particular diagrams to the corresponding anomalous dimensions.
 The coefficients $A,B,C,A'$ and $B'$ are arbitrary and their precise values
are not so important.
 When Yukawa couplings  $Y_{i0}$ are small enough, one can
make expansion in each bracket, and the dependence of these coefficients
disappears.
  However, for large couplings, which are of interest for us
because of the fixed points, we have some residual dependence.
 The requirement that the sum of exponents equals 1 follows from a comparison
with RGEs.
 Solutions are close to the exact ones when the brackets are roughly
equal to each other. Apparently, since $F_\tau < F_t \sim F_b$ and $Y_\tau
\leq Y_b \leq Y_t$ one cannot completely satisfy this requirement.
 Our choice of the coefficients  $A,B,C,A'$ and $B'$ is dictated mainly by
simplicity.
 In the following we choose them as
\begin{eqnarray}
\!\! B=A,C=2/3A\! &\rightarrow & A=7/2,\ B=7/2,\ C=7/3,\\
\!\! B'=A'/2 \!  &\rightarrow & A'=21/4,\ B'=21/8 .
\end{eqnarray}
 This gives approximate solutions like
\begin{eqnarray}
  Y_t\!&\approx&\!\!\frac{\displaystyle Y_{t0}E_t}
        {\displaystyle \left[1+\!\frac{7}{2}(Y_{t0}F_t+Y_{b0}F_b)\right]^{2/7}\!\!
                       \left[1+7Y_{t0}F_t\right]^{5/7}},
                                \label{y1} \\
  Y_b\!&\approx&\!\! \frac{\displaystyle Y_{b0}E_b}
        {\displaystyle \left[1+\frac{7}{2}(Y_{t0}F_t+Y_{b0}F_b)\right]^{2/7}\!\!
 \left[1+7Y_{b0}F_b\right]^{2/7}\left[1+\frac{7}{3}(3Y_{b0}F_b+Y_{\tau 0}F_\tau )\right]^{3/7}} \label{y2}
 ,\\
 Y_\tau \!&\approx& \!\!\frac{\displaystyle Y_{\tau 0}E_\tau }
        {\displaystyle \left[1\!+\!\frac{21}{4}Y_{\tau 0}F_\tau \right]^{4/7}\!\!
  \left[1\!+\!\frac{7}{3}(3Y_{b0}F_b\!+\!Y_{\tau 0}F_\tau )\right]^{3/7}} . \label{y3}\,
\end{eqnarray}
Solutions for $A_i$ and $\Sigma_i$  can be obtained by Grassmannian
expansion with the initial conditions
\begin{equation}\label{init}
 A_i(0)=A_0, \ \ \Sigma_i(0)=\Sigma_0 .
\end{equation}
These initial conditions correspond to the so-called universality hypothesis
which we  follow in our numerical illustration for simplicity. However,
one can choose arbitrary initial conditions for the soft terms when needed.
It leads to an obvious modification of the formulae.

One can get also the corresponding solutions for the individual soft masses.
This can be achieved either by Grassmannian expansion of the corresponding
brackets in (\ref{y1})-(\ref{y3}), or by expressing the masses through the
$\Sigma$s in an exact way.  The second way gives a slightly better
agreement  with numerical solutions (see below). We present the
explicit expressions for the soft terms and masses in Appendix A.
\subsection{Numerical analysis}
We start by investigating the precision of approximate solutions
for the Yukawa couplings.
 To estimate the accuracy, we introduce a relative error which is
defined as
\begin{equation}\label{defer}
\varepsilon  = \frac{Y_{approx}-Y_{numeric}}{Y_{numeric}},
\end{equation}
and corresponds to the $M_Z$ scale ($t=66$) at the end of the
integration range. The accuracy for the solutions of soft terms
 is defined in the same way.

 Let us take at the beginning all  three Yukawa couplings to be
equal at the GUT scale and to have their common value $Y_0$ in the
range $(.01\div 25)\alpha_0$. The upper limit is taken in order
not to leave the perturbativity regime.
 We find that for $Y_0\leq \alpha_0$ the approximation errors
 are less than $3\%$ for all $Y$'s.
 While for $Y_t$ it remains smaller than $2\%$ over the whole
range of initial values at the GUT scale, for $Y_b$ the error
increases up to $4\%$ and for $Y_\tau $ up to $14\% $ (for large
values of $Y_0$). It is worth  mentioning that for small $Y_0$
(around $\alpha_{0}/2$ and below) the accuracy is very good
(fractions of per cent or better).

Consider now $Y_{b0}=Y_{\tau 0}=Y_0 \leq 10\alpha_0$ and let the
top Yukawa coupling vary within the limits   $1\leq Y_{t0}/Y_0
\leq 10$ in order to examine the applicability of our formulae.
In this case the accuracy it is spoilt a little bit with increasing
initial values. Namely, the error for $Y_t$ increases up to
$10\%$, and for $Y_b$ and $Y_\tau$ up to $20\%$. However, if one
keeps  $Y_0$ in the range $(1/100 \div 1/2)\alpha_0$ the accuracy
for $Y_t$ remains better than
$3\%$, and  for $Y_b$ and $Y_\tau $ better than $10\%$.
%--------------------------------------------------------------
%%%  Y approximation error
\begin{figure}[t]
\resizebox{0.4\textwidth}{.8\textwidth}
{
  \includegraphics{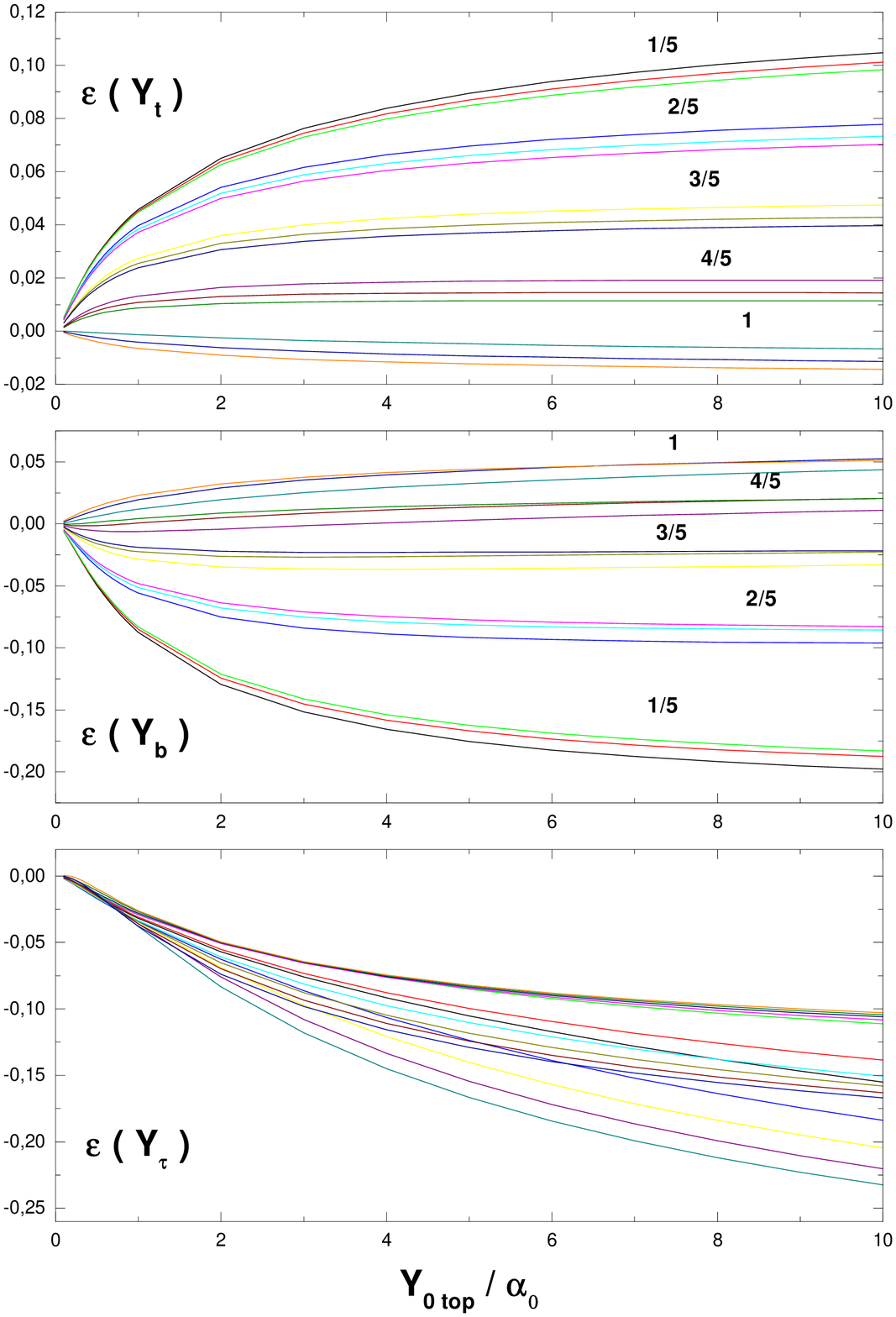}
}
\hspace*{-5pt}
\resizebox{.58\textwidth}{!}
{ 
  \includegraphics{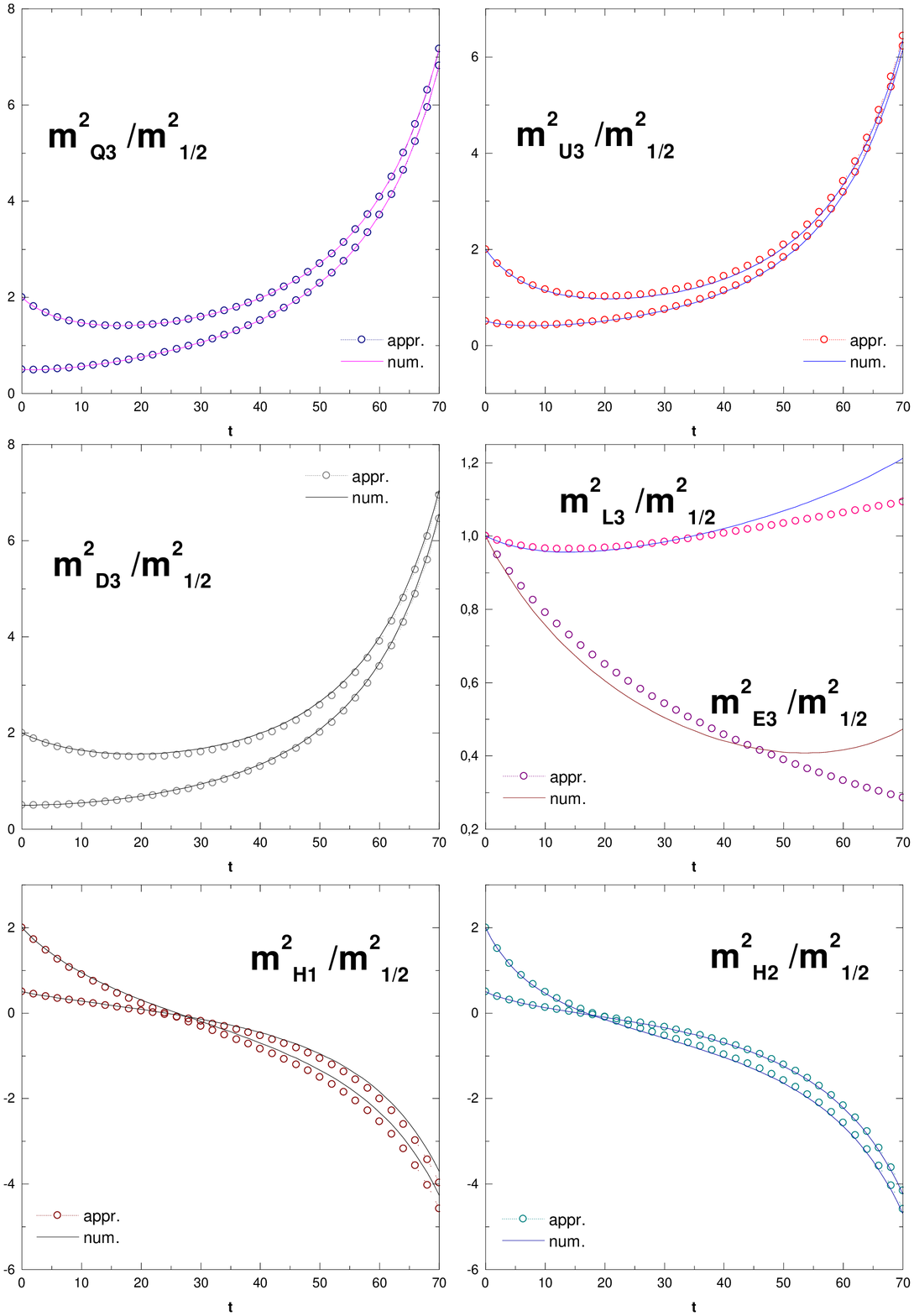}
}
\caption{
{\bf Left side:}
    $Y_t$, $Y_b$ and $Y_\tau$ approximation errors.
     Numerical labels show the ratio
     $Y_{b0}/Y_{t0}$ for the corresponding groups of curves
     (split by different values of $Y_{\tau 0}$).
{\bf Right side:}
 Comparison of approximate and numerical solutions for
  the soft masses. The curves correspond to the following choice of initial
  conditions: $A_{0}=0$, $m^2_0=(1/2)m^2_{1/2}$ and $m^2_0=2m^2_{1/2}$,
  $Y_{t0}=5\alpha_0 ,\, Y_{b0}=Y_{\tau 0}=2\alpha_0$.
     Slepton masses are shown only for $m^2_0=m^2_{1/2}$.
   Dotted lines correspond to analytical approximation,
   solid lines to numerical solution.
 }
\label{errandm}       % Give a unique label
\end{figure}
%--------------------------------------------------------------
The particular case  considered above seems to have the worst
accuracy. This is not surprising since our approximate formulae
are supposed to work best of all when all three Yukawa couplings
are nearly equal. If we  keep $Y_{\tau 0}\leq Y_{b0}\leq Y_{t0}$
and the relative ratios less than $5$, we get an average error of less than
$5\%$ for $Y_t$, about $5\%$ for $Y_b$ and $10\%$ for $Y_\tau $.
This statement is illustrated in Fig.\ref{errandm}. For each  Yukawa
coupling we have plotted the error as a function of $Y_{t0}$ in
the range $(1/10\div 10)\alpha_0$. The ratios are kept within the
region $1\leq Y_{t0}/Y_{b0}\leq 5$ and  $1\leq Y_{b0}/Y_{\tau 0}\leq 3$.

\begin{figure*}
%\vspace{-1.5cm}
\resizebox{.95\hsize}{!}{
  \includegraphics{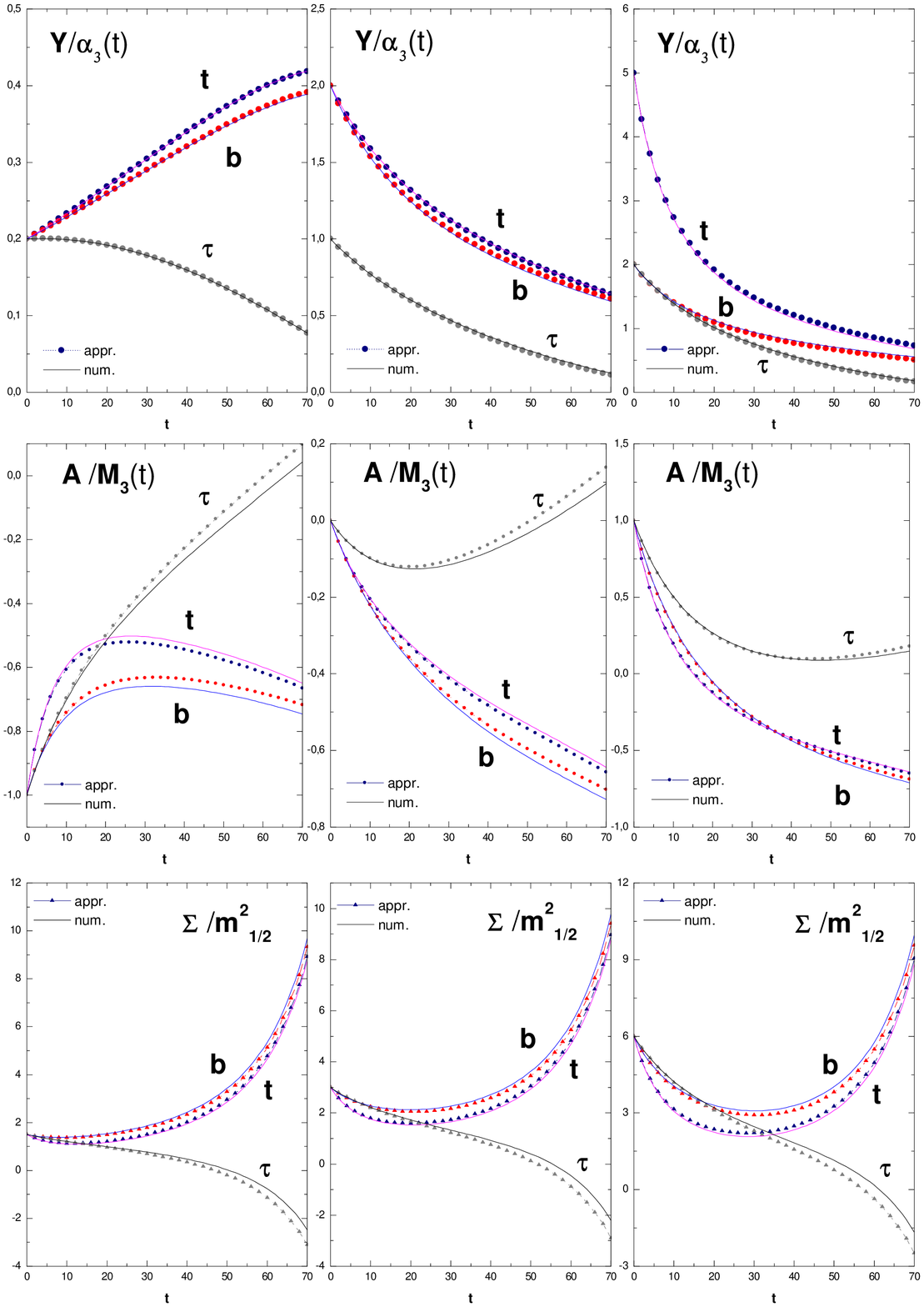}}
\caption{
     Comparison of approximate and numerical solutions for $Y$, $A$ and
    $\Sigma$. Evolution of the soft terms $A$ and $\Sigma$ has been plotted for
    $Y_{t0}=5\alpha_0 \, ,Y_{b0}=Y_{\tau 0}=2\alpha_0$. Dotted lines correspond
    to analytical approximation, solid lines to numerical solution.
                            } \label{evays}
\end{figure*}

Further on, we narrow the range of initial values up to $(1/10\div
10)\alpha_0$ because the errors (defined as in (\ref{defer}))
come to an asymptotic value for $Y_0 > 10\alpha_0$ and almost
vanish for $Y_0 < \alpha_0 /10$. The comparison of numerical and
approximated solutions is  shown in Fig.\ref{evays} for three
different sets of $Y_0$'s. The approximate solutions follow the
numerical ones quite well, preserving their shape, and they have a high
accuracy, especially in the case of equal Yukawa couplings at the GUT
scale. However, as can be seen from the top of Fig. \ref{evays},
one can take  arbitrary initial conditions for the Yukawa couplings,
in particular those which are needed to fit the $t/b/\tau$ masses,
and to use our approximate solutions for these purposes.
%%%%%%%   Y, A, Sigma evolution
%%%%%%%
 For the soft couplings, $A$, we take the initial values at the GUT
scale  to be $A_0=(-2,-1,0,1,2)m_{1/2}$ and leave $Y_0$s in the
narrow range as above. Then, we get an accuracy of $(3\div 5)\%$
for $A_t$ and $A_b$. For $A_{\tau}$ the approximation is worse
when $A_0$ is taken to be negative or smaller than $m_{1/2}$ (see
Fig. \ref{evays}), but things go better for large initial values
of $A_0$ and we get an accuracy of about $10\% $. Again it should
be mentioned that this is an accuracy at the end point where
$A_\tau$ itself is close to 0 and the accuracy defined as
(\ref{defer}) merely gives  an odd hint of the precision. Along the
curves the accuracy is much better. In Fig.\ref{evays} we have
plotted the behaviour of $A_t$, $A_b$ and $A_\tau$ for three
different initial values of $A_0$, namely $\{ -m_{1/2},0,m_{1/2}\}$
and for one set of $Y_0$s.
    As for the $\Sigma$'s, keeping the range of parameter space for
$Y_0$ and $A_0$ as above, we get an accuracy of typically $2\% $
for $\Sigma_{t}$  (even better for fairly equal $Y_0$s). For
$\Sigma_{b}$ the precision is around $4\%$. With $\Sigma_{\tau}$
we get into the same troubles as for $A_{\tau}$. The approximation
becomes good (about $10\%$) only for a large enough ratio of
$m^{2}_{0}/m^{2}_{1/2}$. % and for small initial values of $Y_0$'s.
The approximation errors for $A$'s and $\Sigma$'s are linked with
those for $Y$. If one considers only the sets of small initial
values  for $Y_0$ (less than $\alpha_{0}/2$), then $\Sigma$'s are
approximated with a precision better than $1\%$, regardless of the
$A_0$ values. The precision for $\Sigma$  increases with
$A_0$, but this dependence is not so striking as the one on $Y_0$.

 The approximate formulae for the soft masses may be derived from
$\Sigma$ using (\ref{A1})-(\ref{A7}).
 In this case the approximate
solutions give an accuracy of about $1\div 3\%$ for $m^2_{Q_3}$,
$m^2_{U_3}$ and $(3\div 5)\%$ for $m^2_{D_3}$. For the Higgs
masses we get a good approximation (of about $5\%$ on average) for
$m^2_{H_2}$, and a satisfactory one for  $m^2_{H_1}$ (typically
$10\%$). This accuracy is almost insensitive to the $A_0$
variation (we took it to be in the range $(-2\div 2)m_{1/2}$) and on
the ratio $m^2_{0}/m^2_{1/2}$ (taken to be $0.5\div 2$).
The slepton masses (see Fig. \ref{errandm}) are  not approximated
properly in an analogous way. This is mainly due to the less accurate
approximation of $Y_{\tau}$.

As a concluding remark on numerical analysis, it should be
mentioned that one has a rather good approximation for small (less
than $\alpha_{0}/2$) initial values of the Yukawa couplings. For
larger values of $Y$'s  one has a  good
approximation especially in the case of unification of the Yukawa couplings.
 \subsection{The fixed points}
One can easily see that solutions (\ref{y1})-(\ref{y3}) exhibit the
quasi-fixed point behaviour when the initial values  $Y_{t0}=
Y_{b0}= Y_{\tau 0} \ge \alpha_0$. In this case, one can drop 1 in
the denominator and the resulting expressions become independent
of the initial conditions
\begin{eqnarray}
 Y_t^{FP}&\approx&\frac{E_t}{\displaystyle
        \left[\frac{7}{2}(F_t+F_b)\right]^{2/7}\!\!
        \left[7 F_t\right]^{5/7}} \label{yfp1},\\
 Y_b^{FP}&\approx&\frac{\displaystyle E_b}{\displaystyle
        \left[\frac{7}{2}(F_t+F_b)\right]^{2/7}\!\!\left[7F_b\right]^{2/7}\!\!
        \left[\frac{7}{3}(3F_b+ F_\tau )\right]^{3/7}} \label{yfp2},\\
 Y_\tau^{FP} &\approx& \frac{\displaystyle E_\tau }{\displaystyle
        \left[\frac{21}{4}F_\tau \right]^{4/7}\!\!
        \left[\frac{7}{3}(3F_b+F_\tau )\right]^{3/7}}. \label{yfp3}
\end{eqnarray}
These expressions being expanded over the Grass\-ma\-nni\-an variables
give the quasi-fixed points for the soft terms and masses. The
explicit expressions are presented in Appendix B.
%%%%% FP behaviour
\begin{figure*}
    %\vspace{-1.5cm}
\resizebox{.95\hsize}{!}{
  \includegraphics{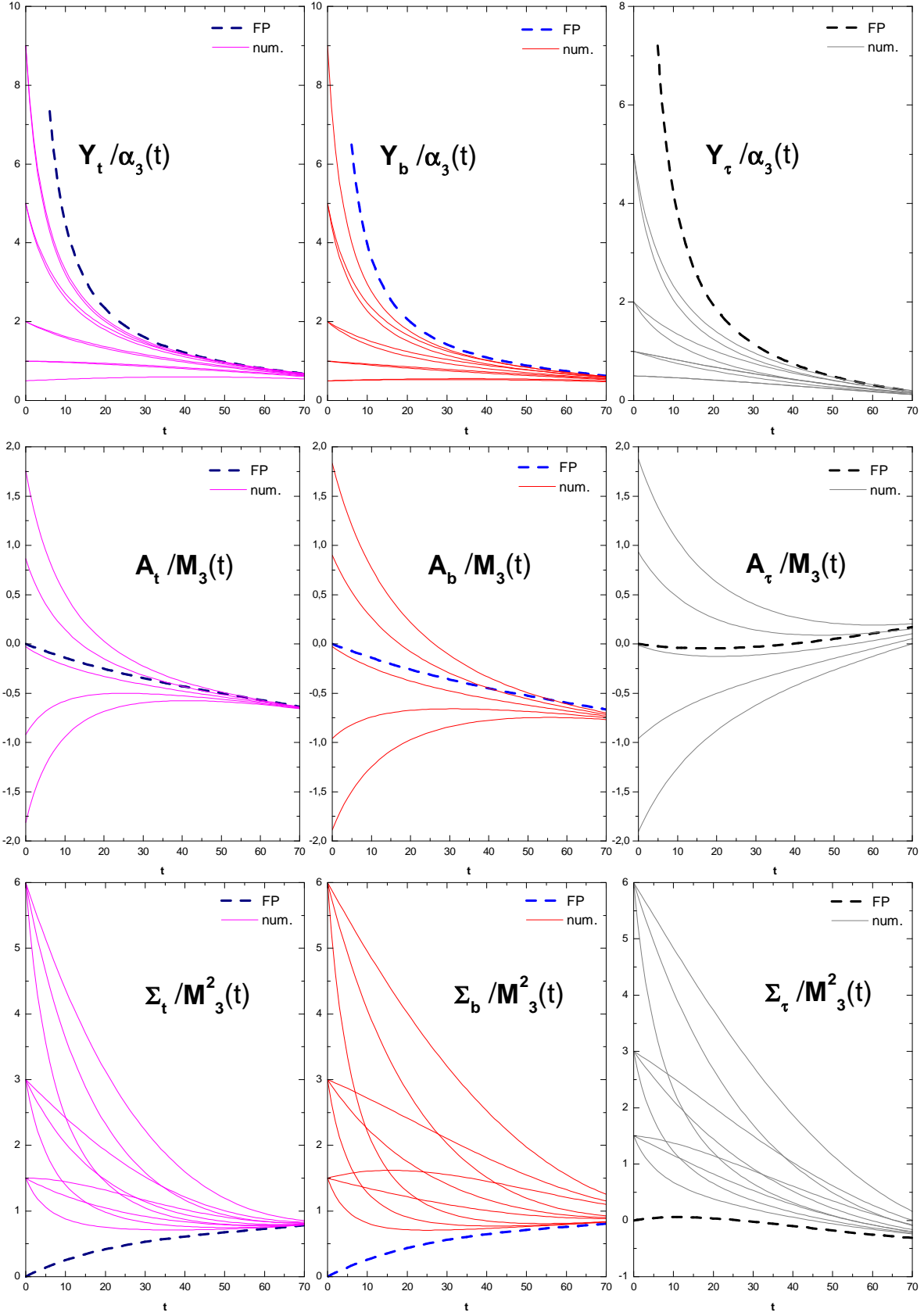}}
 \caption{
     The IRQFP behaviour for Yukawa couplings, soft terms $A_{t,b,\tau}$,and
     $\Sigma$'s. Numerical solutions for all $A$'s are done for
     $Y_{t0}=5\alpha_0$, $Y_{b0}=Y_{\tau 0}=2\alpha_0$. For $\Sigma$ we took $A_0=0$,
     three different sets of  $Y_0$'s and  $m^2_0$. Dotted lines are the IRQFP
     from eqs.(\ref{yfp1}-\ref{yfp3}) and Appendix B.
  }
     \label{fpays}
\end{figure*}
%%%%%%%
We see that the IRQFP behaviour is sharply expressed for $Y_t$ and
$Y_b$ (see Fig. \ref{fpays}), and our approximate solution describes
the fixed point line well. The same takes place for the
corresponding $A$s and $\Sigma$s. For $Y_\tau, A_\tau$ and
$\Sigma_\tau$ the accuracy is worse, however, the solution is
still reliable.
  The soft mass terms there exhibit the same IRQFP behaviour, though
some residual dependence on the initial conditions is left in full
analogy with the exact solutions in the low $\tan\beta$ case. The
approximate solutions allow one to calculate the IRQFP analytically.

  One can see that the fixed points for the soft terms naturally
follow from the Grassmannian expansion of our approximate
solutions (\ref{yfp1})-(\ref{yfp3}) and they inherit their stability
properties, as has been shown in \cite{Jones}. In particular,
the behaviour of $\Sigma$s essentially repeats that of the Yukawa
couplings in agreement with \cite{Z}.

The  existence of the IRQFPs allows one to make predictions for the soft
masses without exact knowledge of the initial conditions. This property
has been widely used (see, for example, \cite{Carena}) and though the
IRQFPs give a slightly larger top mass when imposing $b-\tau$ unification,
it is still possible to fit the quark masses within the error-bars and to
make predictions for the Higgs and sparticle spectrum \cite{JK}.
 This explains general interest in the IRQFPs.
    \section{Discussion}\label{disc}
We hope to convince the reader that the approximate solutions presented
above reproduce the behaviour of the Yukawa couplings with good
precision in the whole integration region and for a large range of
initial values. Relative accuracy is typically a few per cent and is worse only
at the end of the integration region mainly due to the smallness of the
quantities themselves. Moreover, we have shown how the approximate
solutions for the soft terms and  masses follow from those for the
rigid  couplings. This demonstrates how the Grassmannian expansion,
advocated in \cite{K}, works in the case of approximate solutions as well.

  For illustration we have considered universal initial conditions for
the soft terms. In recent time there appeared some interest in non-universal
boundary conditions. Non-universality  can also be included in our
formulae at the expense of changing (\ref{tayl}) and (\ref{init}) using
the same substitution rules, see (\ref{g}) and (\ref{Y}).

Since the form of our approximate solutions has been "guessed" ad hoc
starting from some reasonable arguments, there is no direct way to
improve them. However, one can
imagine more constructive derivation of those solutions which would
allow one to make corrections.
 Needless to say that it is enough to
construct a solution for the rigid terms. Solutions for the soft
terms will follow automatically.
%%%%%%%%%%%%%%%%%%%%%%%
\medskip

{\bf\large Acknowledgement}

{\sf We would like to thank A.V.Gladyshev for valuable discussions. Financial
support from RFBR grants \# 99-02-16650 and \# 96-15-96030 is kindly
acknowledged.}

\appendix
\section*{Appendix A}\label{forms}
We here present approximate expressions for the soft couplings and masses
corresponding to (\ref{y1})-(\ref{y3}):
\hspace*{-4cm}
\begin{eqnarray*}\label{aappr}
 M_i &=& \frac{m_{1/2}}{1+b_i\alpha_0 t} , \\
 A_t &\approx& A_0 \left(1-\frac{Y_{t0}F_t+Y_{b0}F_b}
		{1+\frac 72 (Y_{t0}F_t+Y_{b0}F_b )}
                -\frac{5Y_{t0}F_t}{1+ 7Y_{t0}F_t} \right) \\
         &-& m_{1/2} \left( \frac{t}{E_t}\frac{dE_t}{dt}
                 -\frac{Y_{t0}(tE_t-F_t)+Y_{b0}(tE_b-F_b)}
		{1+\frac 72 (Y_{t0}F_t+Y_{b0}F_b )}
      -\frac{5Y_{t0}(tE_t-F_t)}{1+ 7Y_{t0}F_t} \right)\\
 A_b &\approx& A_0 \left(1-\frac{Y_{t0}F_t+Y_{b0}F_b}
		{1+\frac 72 (Y_{t0}F_t+Y_{b0}F_b )}
                 -\frac{2Y_{b0}F_b}{1+ 7Y_{b0}F_b}
      		-\frac{3Y_{b0}F_b+Y_{\tau 0}F_\tau}
		{1+\frac 73 (3Y_{b0}F_b+Y_{\tau 0}F_\tau )} \right)  \\
         &-&  m_{1/2}\left( \frac{t}{E_b}  \frac{dE_b}{dt}
             -\frac{Y_{t0}(tE_t-F_t)+Y_{b0}(tE_b-F_b)}
		{1+\frac 72 (Y_{t0}F_t+Y_{b0}F_b)} \right.  \\
         &-& \!  \left.  \frac{2Y_{b0}(tE_b-\!F_b)}{1+ 7Y_{b0}F_b}
     -\!\frac{3Y_{b0}(tE_b-\!F_b)+Y_{\tau 0}(tE_\tau-\!F_\tau)}
		{1+\frac 73 (3Y_{b0}F_b+Y_{\tau 0}F_\tau )} \right)\\
 A_\tau &\approx& A_0 \left(1-\frac{3Y_{\tau 0}F_\tau}
		{1+\frac{21}{4} Y_{\tau 0}F_\tau}
                -\frac{3Y_{b0}F_b+Y_{\tau 0}F_\tau}
		{1+\frac 73 (3Y_{b0}F_b+Y_{\tau 0}F_\tau )} \right) \\
         & - & m_{1/2} \left( \frac{t}{E_\tau} \frac{dE_\tau}{dt}
                -\frac{3Y_{\tau 0}(tE_\tau-F_\tau)}
		{1+\frac{21}{4}Y_{\tau 0}F_\tau}
  		-\frac{3Y_{b0}(tE_b-F_b)+Y_{\tau 0}(tE_\tau -F_\tau)}
                {1+\frac 73 (3Y_{b0}F_b+Y_{\tau 0}F_\tau )}  \right)\\
\Sigma_t&\approx &\Sigma_0\left( 1-\frac{Y_{t0}F_t+Y_{b0}F_b}
		{1+\frac 72 (Y_{t0}F_t+Y_{b0}F_b )}
         -\frac{5Y_{t0}F_t}{1+ 7Y_{t0}F_t} \right) \\
&&\hspace*{-2cm} -
	\frac{A_0^2Y_{t0}F_t\!-\!2A_0Y_{t0}
		m_{1/2}(tE_t\!-\!F_t)\!+\!Y_{t0}m_{1/2}^2t^2\frac{dE_t}{dt}
	\!+\!A_0^2Y_{b0}F_b\!-\!2A_0Y_{b0}
	m_{1/2}(tE_b\!-\!F_b)\!+\!Y_{b0}m_{1/2}^2t^2\frac{dE_b}{dt}}
	{1\!+\!\frac72 (Y_{t0}F_t\!+\!Y_{b0}F_b )}\\
&-& \frac{5}{1+7Y_{t0}F_t}\left( A_0^2Y_{t0}F_t 
   -2A_0Y_{t0}m_{1/2}(tE_t-F_t)+Y_{t0}m_{1/2}^2t^2\frac{dE_t}{dt} \right)\\
&+&\!\!\frac{\frac 72}
	{\bigl(1\!+\!\frac 72 (Y_{t0}F_t\!+\!Y_{b0}F_b )\bigr)^2}
       \!\Bigl(-A_0Y_{t0}F_t\!+\!Y_{t0}m_{1/2}(tE_t\!-\!F_t)\Bigr.
         \Bigl.-A_0Y_{b0}F_b+\!Y_{b0}m_{1/2}(tE_b-F_b)\Bigr)^2\!\!\\
& &   \!+\!m_{1/2}^2\frac{d}{dt}\!\left( \frac{t^2}{E_t}\frac{dE_t}{dt}
 	\right)  +  \frac{35}{\bigl(1+7 Y_{t0}F_t\bigr)^2}
	\Bigl(-A_0Y_{t0}F_t+Y_{t0}m_{1/2}(tE_t-F_t)\Bigr)^2 \\
\Sigma_b&\approx &\Sigma_0\left( 1-\frac{Y_{t0}F_t+Y_{b0}F_b}
	{1+\frac 72 (Y_{t0}F_t+Y_{b0}F_b )}
          -\frac{2Y_{b0}F_b}{1+ 7Y_{b0}F_b} \
           -\frac{3Y_{b0}F_b+Y_{\tau 0}F_\tau}{1+
	\frac 73 (3Y_{b0}F_b+Y_{\tau 0}F_\tau )} \right)\\
 &-&\!\frac{1}{1\!+\!\frac 72 (Y_{t0}F_t+\!Y_{b0}F_b )}
            \Bigl(A_0^2Y_{t0}F_t-\!2A_0Y_{t0}m_{1/2}(tE_t\!-\!F_t)\Bigr. \\
 & &\Bigl. +Y_{t0}m_{1/2}^2t^2\frac{dE_t}{dt}
        +\!A_0^2Y_{b0}F_b-2A_0Y_{b0}m_{1/2}(tE_b-F_b)
     +Y_{b0}m_{1/2}^2t^2\frac{dE_b}{dt}\Bigr)\\
 & &    -\frac{2}{1+7Y_{b0}F_b}\left(A_0^2Y_{b0}F_b
  -2A_0Y_{b0}m_{1/2}(tE_b-F_b)+Y_{b0}m_{1/2}^2t^2\frac{dE_b}{dt}\right) \\
&-&\frac{1}{1+\frac 73 (3Y_{b0}F_b+Y_{\tau 0}F_\tau  )}
  \Bigl(3A_0^2Y_{b0}F_b+A_0^2Y_{\tau 0}F_\tau 
	-6A_0Y_{b0}m_{1/2}(tE_b-F_b) \Bigr. \\
& &\Bigl.-2A_0Y_{\tau 0}m_{1/2}(tE_\tau -F_\tau )
          +3Y_{b0}m_{1/2}^2t^2\frac{dE_b}{dt}
          +Y_{\tau 0}m_{1/2}^2t^2\frac{dE_\tau }{dt}\Bigr)\\
&+&\!\!\frac{{\frac{7}{2}}
    \! \Bigl(-A_0Y_{t0}F_t\!+\!Y_{t0}m_{1/2}(tE_t\!-\!F_t)
        -A_0Y_{b0}F_b+\!Y_{b0}m_{1/2}(tE_b-F_b)\Bigr)^2}
	{\bigl(1\!+\!\frac 72 (Y_{t0}F_t\!+\!Y_{b0}F_b )\bigr)^2} \\
& & \!\!+\!m_{1/2}^2\frac{d}{dt}\!\left( 
	\frac{t^2}{E_b}\frac{dE_b}{dt}\right)
  +14 \frac{\bigl(-A_0Y_{b0}F_b+Y_{b0}m_{1/2}(tE_b-F_b)\bigr)^2}
	{\bigl(1+7 Y_{b0}F_b\bigr)^2} \\
&+&\! \frac{{\frac{7}{3}} 
	\Bigl(-3A_0Y_{b0}F_b-\!A_0Y_{\tau 0}F_\tau
         +3Y_{b0}m_{1/2}(tE_b-F_b) 
	+Y_{\tau 0}m_{1/2}(tE_\tau -F_\tau)\Bigr)^2}
	{\bigl(1\!+\!\frac 73 (3Y_{b0}F_b\!+\!Y_{\tau 0}F_\tau)\bigr)^2} \\
 \Sigma_\tau &\approx &
       \Sigma_0\left( 1-\frac{3Y_{\tau 0}F_\tau}{1
	+\frac{21}{4}Y_{\tau0}F_\tau }
 -\frac{3Y_{b0}F_b +Y_{\tau 0}F_\tau}{1+ \frac 73(3Y_{b0}F_b
	+Y_{\tau 0}F_\tau)} \right)\\
  &-& \frac{3}{1+\frac{21}{4} Y_{\tau 0}F_\tau}
      \Bigl(A_0^2Y_{\tau 0}F_\tau -2A_0Y_{\tau 0}m_{1/2}(tE_\tau -F_\tau)
             +Y_{\tau 0}m_{1/2}^2t^2\frac{dE_\tau }{dt}\Bigr)  \\
       &-&\!\!\frac{1}{1+\frac 73 (3Y_{b0}F_b+Y_{\tau 0}F_\tau  )}
            \Bigl(3A_0^2Y_{b0}F_b\!+\!A_0^2Y_{\tau 0}F_\tau
        \!-\!6A_0Y_{b0}m_{1/2}(tE_b\!-\!F_b)\!+\!3Y_{b0}
		m_{1/2}^2t^2\frac{dE_b}{dt}\Bigr.\\
  & &\Bigl. -2A_0Y_{\tau 0}m_{1/2}(tE_\tau -F_\tau )
	+Y_{\tau 0}m_{1/2}^2t^2\frac{dE_\tau }{dt}\Bigr) 
       +\frac{63}{4} \frac{\bigl(-A_0Y_{\tau 0}F_\tau 
	+Y_{\tau 0}m_{1/2}(tE_\tau -F_\tau)\bigr)^2}
            {\bigl(1+\frac{21}{4} Y_{\tau 0}F_\tau \bigr)^2} \\
       &+& \frac{\frac{7}{3}
	\bigl(-\!3A_0Y_{b0}F_b-\!A_0Y_{\tau 0}F_\tau         
        +3Y_{b0}m_{1/2}(tE_b-F_b)+Y_{\tau 0}m_{1/2}(tE_\tau -F_\tau)\bigr)^2}
	{\bigl(1+\frac 73 (3Y_{b0}F_b+Y_{\tau 0}F_\tau )\bigr)^2}\\
       & & \qquad \qquad +m_{1/2}^2\frac{d}{dt}
	\left( \frac{t^2}{E_\tau }\frac{dE_\tau}{dt} \right).\\
\end{eqnarray*}
To find the individual soft masses one can formally perform integration of
the RG equations  and express the masses through $\Sigma$s solving a
system of linear algebraic equations. This gives
\begin{eqnarray}
m_{Q_3}^2\!\!&=&\!\!\frac{13}{61}m^2_0+
	m_{1/2}^{2}\left(\frac{64}{61}f_{3}+\frac{87}{122}f_{2}
                        -\frac{11}{122}f_{1}\right)\!\! 
             +\frac{1}{122}\left(17\Sigma_t
	+20\Sigma_b-5\Sigma_{\tau}\right)  \label{A1}\\
m_{U_3}^2\!\!&=&\!\!\frac{7}{61}m^2_0+m_{1/2}^{2}
	\left(\frac{72}{61}f_{3}-\frac{54}{61}f_{2}+
                        \frac{72}{305}f_{1}\right)\!\!
           +\frac{1}{122}\left(42\Sigma_t-8\Sigma_b+2\Sigma_{\tau}\right) \\
m_{D_3}^2\!\!&=&\!\!\frac{19}{61}m^2_0+m_{1/2}^{2}
	\left(\frac{56}{61}f_{3}-\frac{42}{61}f_{2}+
                         \frac{56}{305}f_{1}\right)\!\!
    +\frac{1}{122}\left(-8\Sigma_t+48\Sigma_b-12\Sigma_{\tau}\right) \\
m_{H_1}^2\!\!&=\!\!&-\frac{32}{61}m^2_0+m_{1/2}^{2}
	\left(-\frac{120}{61}f_{3}\!-\frac{3}{122}f_{2}\!-
                        \frac{57}{610}f_{1}\right)\!\!
  +\frac{1}{122}\left(-9\Sigma_t\!+54\Sigma_b\!+17\Sigma_{\tau}\right) \\
m_{H_2}^2\!\!&=&\!\!-\frac{20}{61}m^2_0+m_{1/2}^{2}
		\left(-\frac{136}{61}f_{3}+\frac{21}{122}f_{2}-
                                 \frac{89}{610}f_{1}\right)\!\!
      +\frac{1}{122}\left(63\Sigma_t-12\Sigma_b+3\Sigma_{\tau}\right) \\
m_{L_3}^2\!\!&=&\!\!\frac{31}{61}m^2_0+m_{1/2}^{2}
		\left(\frac{40}{61}f_{3}+\frac{123}{122}f_{2}-
                                 \frac{103}{610}f_{1}\right)\!\!
      +\frac{1}{122}\left(3\Sigma_t-18\Sigma_b+35\Sigma_{\tau}\right) \\
m_{E_3}^2\!\!&=&\!\!\frac{1}{61}m^2_0+m_{1/2}^{2}
	\left(\frac{80}{61}f_{3}-\frac{60}{61}f_{2}+
         \frac{16}{61}f_{1}\right)\!\! \label{A7}
             +\frac{1}{122}\left(6\Sigma_t-36\Sigma_b+70\Sigma_{\tau}\right).
\end{eqnarray}
The masses of squarks and sleptons of the first two generations are given by
(\ref{me})-(\ref{mq}).
%%------------------------------------------
\section*{Appendix B}\label{forms2}
We present here the IRQFPs for the soft couplings and masses. They are obtained via
Grassmannian expansion of (\ref{yfp1})-(\ref{yfp3}).
\begin{eqnarray*}\label{fpsoft}
 A_t^{FP}\!\!\!\!&\approx&\!\!\!\!- m_{1/2} \left(\frac{t}{E_t}
	  \frac{dE_t}{dt}\!
     -2\frac{(tE_t\!-F_t)\!+(tE_b\!-F_b)}{ 7 (F_t\!+F_b)}
	-\frac{5(tE_t\!-F_t)}{ 7F_t} \right)\\
 A_b^{FP} \!\!\!\!&\approx&\!\!\!\!
-  m_{1/2}\!\left(\!\frac{t}{E_b}\!\frac{dE_b}{dt}\!
	-\!2\frac{(tE_t\!-\!F_t)\!+\!(tE_b\!-\!F_b)}{7 (F_t\!+\!F_b )}
   \!-\frac{2(tE_b\!-\!F_b)}{ 7F_b} 
       \!-3\frac{3(tE_b\!-\!F_b)\!+\!(tE_\tau\!
	-\!F_\tau)}{7(3F_b\!+\!F_\tau )}\!\right)\\
 A_\tau^{FP} \!\!\!\!&\approx&\!\!\!\!-  m_{1/2} 
	\left( \frac{t}{E_\tau}\frac{dE_\tau}{dt}
  \!-\frac{4(tE_\tau\!-F_\tau)}{7F_\tau}\! -3\frac{3(tE_b\!-F_b)+(tE_\tau\!
 -F_\tau)}{7 (3F_b\!+F_\tau )}\right)\\
\Sigma_t^{FP}\!&\approx &\! m_{1/2}^2\left(-\frac 27 \
	 \frac{t^2\frac{dE_t}{dt}
        +t^2\frac{dE_b}{dt}}{ (F_t+F_b )} -\frac 57\
        \frac{t^2\frac{dE_t}{dt}}{F_t}+ \frac{d}{dt}\left(
        \frac{t^2}{E_t}\frac{dE_t}{dt} \right)\right.\\
     & &+\left. \frac 27\
        \frac{[(tE_t-F_t)+(tE_b-F_b)]^2}{ (F_t+F_b )^2}  + \frac 57\
        \frac{(tE_t-F_t)^2}{ F_t^2} \right)\\
 \Sigma_b^{FP}\!&\approx &\! m_{1/2}^2\left(-\! \frac 27\!\
    \frac{t^2\frac{dE_t}{dt}\!+\!t^2\frac{dE_b}{dt}}
	{(F_t\!+\!F_b )}\!-\!\frac 27\! \
    \frac{t^2\frac{dE_b}{dt}}{F_b}\! -\! \frac 37 \ \frac{3t^2\frac{dE_b}{dt}
    \!+\!t^2\frac{dE_\tau }{dt}}{ (3F_b\!+\!F_\tau )}\right. \\
        & &\left. +  \frac 27 \
    \frac{[(tE_t-F_t)+(tE_b-F_b)]^2}{ (F_t+F_b )^2}  + \frac 27\
    \frac{(tE_b-F_b)^2}{ F_b^2} \right. \\
    & &\left.  + \frac 37 \
    \frac{[3(tE_b-F_b)+(tE_\tau -F_\tau)]^2}{ (3F_b+ F_\tau )^2} +
     \frac{d}{dt}\left( \frac{t^2}{E_b}\frac{dE_b}{dt}\right)\right) \\
 \Sigma_\tau^{FP} \!&\approx&\! m_{1/2}^2\!\left( -\frac 47\!\
        \frac{t^2\frac{dE_\tau }{dt} }{F_\tau}
         -\frac 37\!\ \frac{3t^2\frac{dE_b}{dt} +t^2\frac{dE_\tau}{dt}}
        {(3F_b+\!F_\tau )} +\! \frac{d}{dt}\!\left( \frac{t^2}{E_\tau}
        \frac{dE_\tau}{dt} \right) \right. \\
     & &\left.  +\frac 47\! \frac{(tE_\tau -F_\tau)^2}{F_\tau^2} +
         \frac 37\ \frac{[3(tE_b-F_b)+(tE_\tau -F_\tau)]^2}{(3F_b+F_\tau)^2} 
	\right)\\
\end{eqnarray*}
% ----------------------------------------------------------------

\end{document}